\documentstyle[epsfig,aps]{revtex}

\begin{document}

\author{H. D. Sivak$^{1}$, A. P\'{e}rez$^{3}$ and Joaquin Diaz-Alonso$^{1,2}$}
\address{$^{1}$D.A.R.C., Observatoire de Paris-Meudon\\
92195 Meudon (France)\\
$^{2}$Departamento de F\'{\i}sica \\
Facultad de Ciencias \\
Universidad de Oviedo. \\
33001 Oviedo (Spain)}
\address{$^{3}$Departamento de F\'{\i}sica Te\'{o}rica\\
Universidad de Valencia. \\
46100 Burjassot (Valencia) Spain}
\title{Complex poles and oscillatory screened potential in QED plasmas}

\maketitle

\begin{abstract}
We discuss the mechanism for the oscillatory behavior of the static
interparticle potential in a degenerate electron plasma. This behavior,
observed in  metallic alloys, is commonly referred to as 'Friedel
oscillations', and its origin associated to the Kohn singularity. We show
that, although this interpretation is adequate for large distances, the
oscillations at short distances originate from a complex pole of the in-medium
photon propagator in the complex q-plane, which exists aside the (purely
imaginary) Debye pole. Such short-range oscillations can be physically
discriminated if they remain at temperatures for which Friedel oscillations
have already disappeared. This is suggested by finite temperature calculations
in non-electromagnetic plasma models showing a similar pole structure.
\end{abstract}


It is well-known that wave propagation and particle interaction in vacuum are
substantially modified by medium effects inside a plasma at a given density
and temperature \cite{FetterWal,KA89,LeBellac}. In the case
of a QED plasma, ordinary transverse photons (with respect to the wavelength
vector) acquire an 'effective mass', and are governed by a dispersion relation
which substantially differs from the vacuum wave equation. In addition to these
photons, there appear also longitudinal modes with different propagation properties
\cite{BS93,R96}. These modes arise from the poles of the photon propagator
in the medium, and give raise to a number of phenomena in plasma and solid state
physics. It is then important to analyze the structure of the photon propagator
and its physical consequences. In this letter, we will concentrate on this study
for the static photon propagator at zero temperature in the complex q-plane
(where q is the momentum of the photon), and examine some of the consequences
of our findings when considering the interparticle potential in the plasma.

We calculate the quasi-photon propagator in the RPA approximation to the dielectric
function, and include a local field correction, as given by Ichimaru \cite{ICH82}
(we have also performed calculations with other parameterizations of this correction).
The expression of the static potential between two point-like impurities of
charge $e$ is obtained from this propagator, and reads :

\begin{equation}
\label{V(r)}
V(r)\, \, =\, \, \frac{e^{2}}{4\pi ^{2}r}{\textit {Im}}\int _{-\infty }^{\infty }dq\frac{qe^{iqr}}{D(q)}
\end{equation}

where \( D(q)=q^{2}\epsilon (\omega =0,q) \), \( \epsilon (\omega ,q) \) is
the dielectric function and \( r \) the interparticle distance.

The denominator \( D(q) \) contains the function \( f(q)=\log |\frac{q+2p_{f}}{q-2p_{f}}| \),
with \( p_{f} \) the Fermi momentum of the electrons. This function can be
written, for real values of \( q \), as follows :

\begin{equation}
\label{extlog}
f(q)=\left\{ \begin{array}{c}
f_{1}(q)\equiv 2\arg \tanh (\frac{q}{2p_{f}}),|\frac{q}{2p_{f}}|<1\\
f_{2}(q)\equiv 2\arg \tanh (\frac{2p_{f}}{q}),|\frac{2p_{f}}{q}|<1
\end{array}\right. 
\end{equation}

We define the functions \( D_{1}(q) \) and \( D_{2}(q) \), which are obtained
by replacing \( f(q)\rightarrow f_{1}(q) \) and \( f(q)\rightarrow f_{2}(q) \),
respectively, in \( D(q) \). 
Then, Eq. (\ref{V(r)}) becomes
\begin{equation}
V (r)\ =\ \frac{e^{2}}{4\pi ^{2}r}
{\it Im}\left[ \int_{-\infty}^{-2p_{f}}dq
 \frac{qe^{iqr}}{D_{2}(q)}+
\int_{2p_{f}}^{\infty }dq \frac{qe^{iqr}}{D_{2}(q)} + 
\int_{-2p_{f}}^{2p_{f} }dq \frac{qe^{iqr}}{D_{1}(q)}
\right] \label{V1(r)}
\end{equation}
We next consider the analytical continuation of the function $D_{1}(q)$ to 
the complex q-plane, and apply the residue theorem 
in order to transform the last 
integration in Eq. (\ref{V1(r)}) into the integration domain of the first two 
integrals in the same equation. After this operation, and with the help of 
the symmetry properties of the photon propagator due to the Onsager relations, 
Eq. (\ref{V(r)}) can be cast under the form :

\begin{equation}
V(r)=V_{P}(r)+V_{C}(r)
\end{equation}

Here, \( V_{P}(r) \) is the contribution of the residues of \( 1/D_{1}(q) \)
in the region \( Im(q)>0 \). We have also defined

\begin{equation}
V_{C}(r)\ =\ \frac{e^{2}}{2\pi ^{2}r}{\it Im}\left[
 \int_{2p_{f}}^{\infty }dq q e^{iqr} \left(
 \frac{1}{D [f_{2} (q)]} - \frac{1}{D [f_{2} (q)+ i \pi] } \right)\right]
\end{equation}

\begin{figure}
\includegraphics[width=12cm]{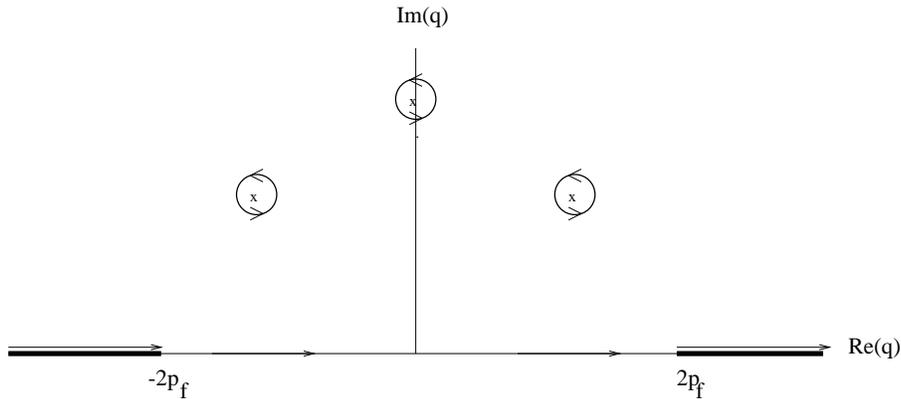}
\caption{Analytical structure of the function $1/D_1(q)$ in the complex $q$-
plane. Crosses indicate poles (discussed in the text). Thick lines
correspond to branch cuts. The integration contour is shown by thin lines with
arrows.}
\label{complexq}
\end{figure}

The analytical structure of \( 1/D_{1}(q) \) on the upper half-plane is shown
in Fig. \ref{complexq}, where we have also drawn the integration contour used
to perform the integration. As we discuss later, in addition to the well-known
'Debye pole', lying on the imaginary axis, \emph{there exist complex poles},
which are responsible for strong oscillations of the interparticle potential
at short distances.

In fact, while the oscillatory behavior of the screened potential created by
a  ionic impurity in metallic alloys is known since the works of Friedel 
\cite{FR52} (see also \cite{FetterWal,KA88}), its algebraic decay with
distance  has been associated to the existence of the  Kohn singularity
\cite{KO59} in the photon self-energy, induced by the sharp  edge of the
degenerate electron distribution: when considering energy and  momentum
conservation in the collisions between the electrons and soft  quasi-photons,
we can easily see that only a fraction of the electrons  can interact with 
quasi-photons and, therefore, screening of a Fermi gas at T=0 K is less
effective  than screening at finite temperature. At the threshold of the
interaction, i. e. at the threshold of electron-hole creation, the momentum
of the  quasi-photon is equal to the diameter of the Fermi sphere. It must be 
noticed that the photon self-energy presents a singularity at this point. The 
presence of a singularity permits to obtain, with the help of Lighthill's 
method, the asymptotic form ($r\longrightarrow \infty $) of the potential,  as
an expansion in terms of the form $\cos (2p_{f}r)$ and $\sin (2p_{f}r)$,
damped as negative powers of $r$, and enhanced by powers of $\log(4 p_{f} r)$ 
\cite{LI64}. One has to remember, for further discussion, that Lighthill's
method  consists essentially in replacing the Fourier (anti) transform of a 
non-analytical function by the (anti) transform of the (formal) expansion  of
that function around its singular points.

\begin{figure}
\includegraphics[width=8cm,angle=270]{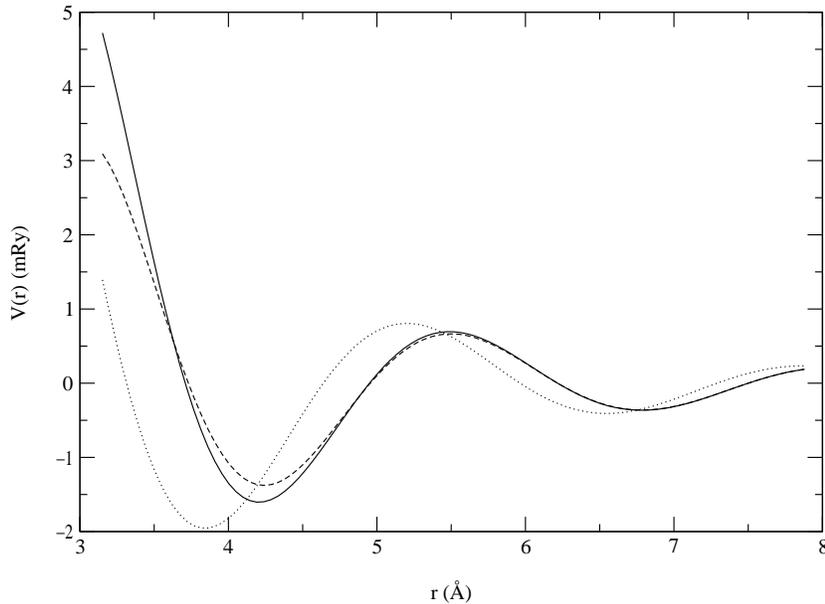}
\caption{Two different approximations to the interparticle potential at large
distances. The dotted line shows the first term of the Lighthill expansion,
and the dashed line corresponds to $V_C(r)$. Also shown for comparison is the
exact result obtained by direct numerical calculation (solid line).}
\label{Lighthill}
\end{figure}

In Fig. \ref{Lighthill} we have plotted the result for the first, dominant 
term ($\sim \cos (2p_{f})/r^{2}$), of Lighthill's expansion (dotted line) for 
the potential around an impurity placed into an electron gas 
with \footnote{This is a typical value corresponding to metals.} $r_{s}=3$, 
where \( r_{s} \) is the characteristic interparticle distance in units of
the Bohr's radius. In the same figure we have also plotted for comparison (solid
line) the result of a numerical calculation of \( V(r) \), as obtained by direct
integration from Eq. (\ref{V(r)}). As we see, the Lighthill expansion converges
to the true result for large values of \( r \). An even better approximation
to the exact result is given by \( V_{C}(r) \), as defined above, and represented
by the dashed line. The reason for this good agreement is that \( V_{P}(r) \)
exponentially goes to zero when \( p_{f}r\gg 1 \), and thus only \( V_{C}(r) \)
is important for long distances. Of course, the advantage of the Lighthill's method
for large distances is to provide a systematic way to construct corrections
to the simple formula we used here, and to give an analytical expression for
these terms, although they become rapidly cumbersome as the order of the approximation
increases.

\begin{figure}
\includegraphics[width=8cm,angle=270]{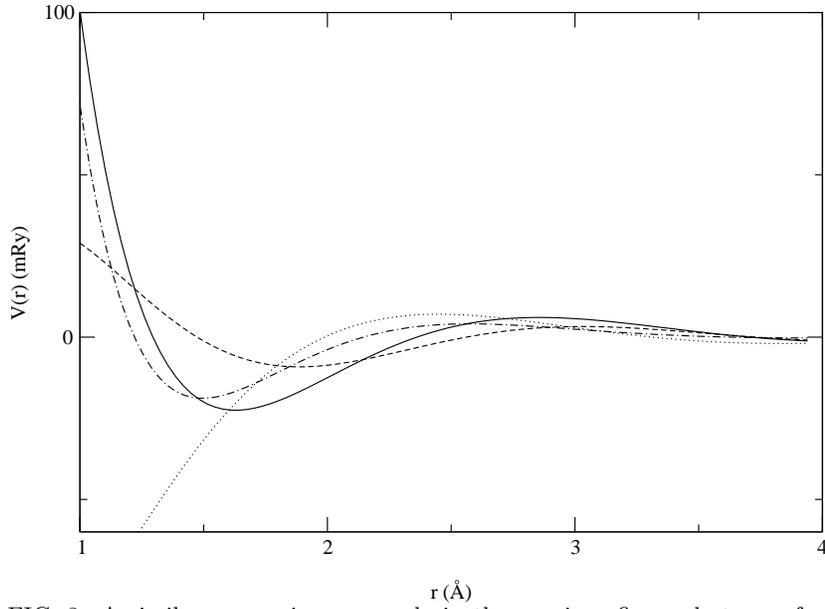}
\caption{A similar comparison as made in the previous figure, but now for  
shorter distances. We have included here the crucial contribution $V_P(r)$ from
the poles (dashed-dotted line).}
\label{shortr}
\end{figure}

However, as we go to shorter distances, the situation becomes quite different.
The reason is that, in this range, the dominant contribution to the form of
the potential is given by the poles of the photon propagator in the complex
plane, which induce exponentially-damped oscillations. This is clearly seen
in Fig. \ref{shortr}, where we have represented the different contributions
to the potential for the same value \( r_{s}=3 \), but for a shorter distance
range. As before, the exact result (obtained from direct numerical integration)
has also been plotted. In order to calculate \( V_{P}(r) \), we have to locate
numerically the zeros of \( D_{1}(q) \) in the upper half-plane, while \( V_{C}(r) \)\ is
calculated by numerical integration. As we already mentioned, we find a zero
lying on the imaginary-\( q \) axis, which is commonly referred to as the 'Debye
pole'. In addition to this, there exists a genuinely complex pole, with both
a real and an imaginary part, which gives raise to exponentially-damped oscillations.
As we can see, the contribution from the poles \( V_{P}(r) \), represented
by the dashed-dotted line, gives a much closer result to the exact line than
the Lighthill term. We have also plotted, for the sake of completeness, the
contribution from \( V_{C}(r) \), which accounts for the difference between
the poles term and the exact potential\footnote{
We have verified that the addition of \( V_{C}(r) \) to \( V_{P}(r) \) reproduces
the exact potential, within machine-size precision numbers.
}. Thus, for this range of distances, the potential is dominated by the pole
contribution. As a matter of fact, for the present value of \( r_{s} \) the
contribution of the Debye and complex poles to \( V_{P}(r) \) are of the same
order.

\begin{figure}
\includegraphics[width=8cm,angle=270]{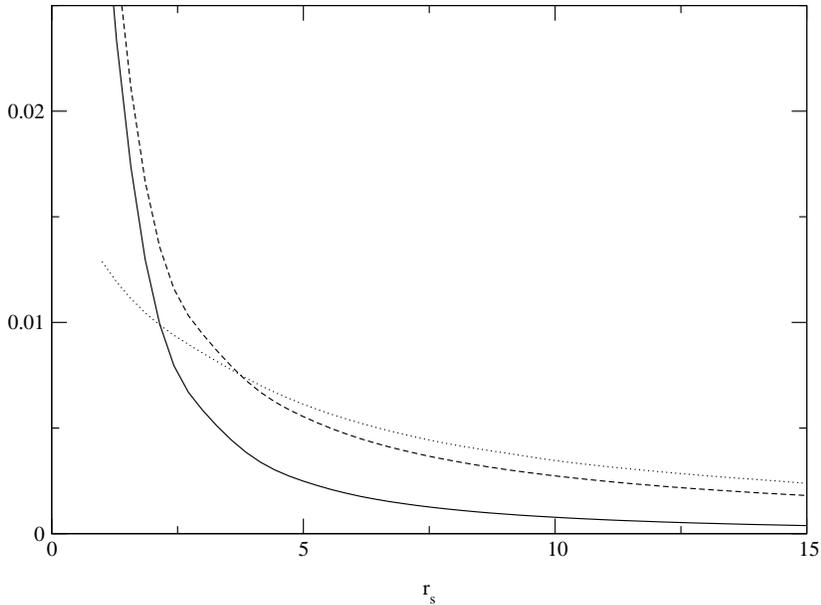}
\caption{Evolution of the real part (dashed line) and the imaginary part (solid
line) of the complex pole as the parameter $r_s$ changes. The dotted line shows
the purely imaginary Debye pole. All magnitudes in ordinates are expressed in
units of the electron mass.}
\label{evolrs}
\end{figure}

As the density is changed, the relative importance of these two contributions 
to \( V_{P}(r) \)
changes. This can be understood from Fig. \ref{evolrs}, where we have shown
the evolution of the real and imaginary parts of the complex pole, and the imaginary
part of the Debye pole. For large densities (\( r_{s}\lesssim 2 \)), \( V_{P}(r) \)
reduces to the well-known Debye screened potential, since the contribution arising
from the complex pole is strongly damped. For larger values of \( r_{s} \),
however, the situation is reversed, since \( V_{P}(r) \) is dominated by the
complex pole and, therefore, this term alone constitutes a good approximation
to the exact potential.

Let us summarize our results. We have studied the static interparticle
potential in a QED electron plasma at zero temperature. This can be obtained
from the knowledge of the dielectric function (or equivalently, of the photon
polarization), which was taken from the RPA approximation, with local field
corrections included. For the latter, we adopted the formulae given by
Ichimaru, although we have performed also calculations with the formulae of
Ref. \cite{FHER93}, with very similar results. We found that the potential
shows an oscillatory behavior for the whole distance range. This behavior has
been traditionally referred to as Friedel oscillations, and its origin
attributed to the Kohn singularity. We have shown that one can get a much
better insight into this behavior through the study of the analytic structure
of the photon propagator in the complex-\( q \) plane. We find that, in
addition to the well-known Debye pole on the imaginary axis, there exists a
complex pole, which accounts for the oscillations of the potential at short
(\( r\gtrsim 1 \AA  \)) distances.

A similar pole structure, and the associated short-range oscillations, has
been found in other (non electromagnetic) plasma models describing the
one-pion exchange \cite{DPS89} and the exchange of other mesons in nuclear
matter \cite{DGP94}. In these cases, one can easily show that the separation
of the screened potential into a short-range component and a Friedel-like
component is not merely a mathematical artifice, but in fact can have
physically observable consequences. Indeed, the phenomenon we have revisited
(Friedel oscillations) and the new one obtained here come from two  different
effects. Friedel oscillations arise from the non-analytical  behavior of the
Fourier transform of the potential created by an impurity,  while our results
come from the zeros of the dispersion relation, in the complex plane, of the
quasi-photon momentum. Using  Maxwell equations, it can be easily obtained the
charge distribution from  the potential of the system. It comes out that the
form of the charge and  potential distribution are qualitatively identical.
The presence of an impurity, then, leads to the apparition of a spatial
static structure of  charges in the medium, which is characteristic of a
highly-interacting system. As long as the temperature increases, the kinetic
energy of particles increases, and this structure breaks down. At the same
time, the non-analytic  behavior of the photon self-energy disappears at
finite temperature, and therefore Friedel oscillations are exponentially
damped with temperature. 

This effect has been explicitly shown in \cite{DPS89} for the one-pion exchange.
We also found that the complex pole is more stable,
and remains present (as well as the associated short-range oscillations), for
higher temperatures. This fact should permit to experimentally distinguish
between both  phenomena.

Even if the immediate extrapolation of these results to the present case might
not be justified, the similarity of the analytic structure of the dressed
boson propagators in both systems at \( T=0 \) suggest that a similar behavior
can be expected for the electronic plasma. If such a thermal behavior is
genuine, it could probably be experimentally tested in heated liquid metals,
although a calculation of the characteristic temperatures at which Friedel and
short-range oscillations disappear will be necessary.

Consequently, in order to get a better understanding of this phenomena and their
possible consequences in plasma physics, it would be interesting to perform
a study similar to the present one at non-zero temperature. Unfortunately, to
do this we need a local-field correction formula valid at finite temperature.
To our knowledge, such formula has not been given yet in the literature.

\textbf{Acknowledgments}

This work has been partially supported by the Spanish Grants DGES PB97-1432 and
AEN99-0692.

\vspace{2cm}

\vspace{2cm}

\vspace{2cm}

\vspace{2cm}

\end{document}